\documentclass[aps,pra,onecolumn,superscriptaddress]{revtex4}

\usepackage{enumerate}

\usepackage{amsmath}
\usepackage{amsfonts}
\usepackage{amssymb}
\usepackage{enumitem}
\usepackage{float}
\usepackage{epsfig}
\usepackage{indentfirst}

\newcommand{\rmd}{{\mathrm d}}
\newcommand{\rme}{{\mathrm e}}

\newcommand{\RR}{{\mathbb R}}

\begin{document}


\def\UnBrasilia{Instituto de F{\'\i}sica,
Universidade de Bras{\'\i}lia, 70910-900 - Bras{\'\i}lia, Brazil}

\def\ICCMPBrasilia{International Center for Condensed Matter Physics,
Universidade de Bras{\'\i}lia, 70910-900 - Bras{\'\i}lia, Brazil}

\def\Marseille{Aix-Marseille Universit\'e, UMR 7345 CNRS,
campus Saint-J\'er\^ome, case 322, \\ 
av.\ esc.\  Normandie-Niemen, FR-13397 Marseille cedex 20, France, EU}

\def\Recife{Departamento de F{\'\i}sica, Universidade Federal Rural de Pernambuco, \\
Rua Manoel de Medeiros, s/n - Dois Irm\~aos, 52171-900 - Recife, Brazil}

\title{Hard-core collisional dynamics in Hamiltonian mean-field model}

\author{I.~Melo}
\affiliation{\Marseille}
\affiliation{\UnBrasilia}

\author{A.~Figueiredo}
\affiliation{\UnBrasilia}
\affiliation{\ICCMPBrasilia}

\author{T.~M.~Rocha Filho}
\affiliation{\UnBrasilia}
\affiliation{\ICCMPBrasilia}

\author{L.~H.~Miranda Filho}
\affiliation{\Recife}

\author{Y.~Elskens}
\affiliation{\Marseille}

%
%

\begin{abstract}
	We consider a modification of the well studied Hamiltonian Mean-Field model with cosine potential by introducing a hard-core point-like repulsive interaction
	and propose a numerical integration scheme to integrate its dynamics. Our results show that the outcome of the initial
	violent relaxation is altered, and also that the phase-diagram is modified with a critical temperature at a higher value than in its
	counterpart without hard-core collisions.
\end{abstract}


\maketitle

\section{Introduction}
\label{sec.Intro}

When performing molecular dynamics simulations of a classical many-particle system, one is usually concerned with two-particle interactions,
characterized by an interaction potential $V(r)$ depending on the distance $r$ between the two particles.
A potential in a $D$-dimensional space is considered long-ranged if it decays at large distances as $1/r^{\alpha}$, with $\alpha<D$,
and conversely short-ranged if $\alpha>D$~\cite{newbook,proc1,proc2,proc3,AC}. While systems with short-range interaction reach the
thermodynamic equilibrium with a relatively short relaxation time and Gaussian final state depending only on the energy
(and possibly other conserved quantities such as linear and angular momentum), systems with long-range interactions have a richer phenomenology.
Among unusual properties that are observed in the latter, we may cite non-additivity of the thermodynamic functions, very long relaxation times to
equilibrium diverging with the number of particles~\cite{NEQ, scaling,scaling2, NEQ2, NEQ3, NEQ4}, negative heat-capacity~\cite{AC,RM}
and non-ergodicity~\cite{ERG, ERG1, ERG2, ERG3},
although non-additive extensivity can be recovered
using Kac's prescription~\cite{kac}. Since a negative heat-capacity is impossible in the canonical ensemble, but is observed in many instances of
isolated (not coupled to a thermal bath) long-range systems, ensemble inequivalence is also possible.
Examples of systems with a long-range interaction potential are:
self-gravitating systems~\cite{SGS,S95,chavanis}, plasmas~\cite{K90, K92, PLP,PLP1}, 
turbulence in two dimensions, cold Coulomb systems and dipolar systems~\cite{newbook},
and different models such as one-dimensional gravity (sheets model)~\cite{MFG, MFG1, MFG2}, 
the self-gravitating Ring Model~\cite{RM, RM1}
and the simple cosine Hamiltonian Mean Field model (HMF)~\cite{HMF}. 
The latter has been extensively studied in the literature due to its simplicity, for being
solvable at equilibrium and allowing much faster molecular dynamics simulations.

The dynamical process in the long-range systems out of equilibrium comprises a short transient violent relaxation into a
quasi-stationary state (QSS) \cite{LB, LB1} with a long lifetime diverging with the number of particles $N$.
Contrary to the thermodynamic equilibrium, the QSS depends strongly on initial conditions~\cite{LTB,LTB1,chavanis2}.
In fact, it depends on an infinite number of conserved quantities (see~\cite{jphyarocha} for a more detailed discussion on this point).
In the present work, we are interested in the study of the dynamics of mixed systems, i.e.\ systems with a long-range potential but also with a
strong core interaction at short distances. Previous works in this direction include the Ising model with neighbor interactions~\cite{ISM, ISM1, ISM2}
and the HMF model modified by adding a short-range term~\cite{HMFM, HMFM1, HMFM2}. In both cases, the results obtained are similar to the
strict long-range case. Here we will consider a hard-core interaction at zero distance. For particles of equal mass, this potential has simply the
effect of interchanging the momenta, which is e\-qui\-val\-ent to a simple label swapping between the two particles involved. As a clear
consequence, this has no effect whatsoever on the evolution of the one-particle distribution function, i.e.\ it does not affect the corresponding
kinetic equation. This is no longer the case for an $N$-particle system with different masses.

We consider in the present work a modified version of the HMF model by introducing a zero-distance hard-core potential and
different masses, and assess its effect on the dynamics of the system, the relaxation to the final thermodynamic
equilibrium, and mass segregation in the HMF model~\cite{zolacir}.
The paper is structured as follows: The model is described in section~\ref{sec.HMF}, and the simulation algorithm in section~\ref{secMD}.
Section~\ref{sec.rslts} presents our results, and we conclude and outline some perspectives in section~\ref{sec.cncl}.

\section{The model}
\label{sec.HMF}

The HMF model is composed of $N$ particles on a unit circle with Hamiltonian
\begin{eqnarray}
  H 
  & = & K(p) + V(\theta) 
  \nonumber  \\ 
  & = &  \sum_{i=1}^{N}\frac{p_i^2}{2m_i}+\frac{1}{2N}\sum_{i,j=1}^{N}\left[1-\cos\left(\theta_{i}-\theta_{j}\right)\right],
	\label{eq:HMF}
\end{eqnarray}
where $m_i$ is the mass, $\theta_{i}$ and $p_i$ are the angular position and conjugate momentum  of the $i$-th particle, respectively, 
$K$ is the total kinetic energy and $V$ the potential energy.
The prefactor $1/N$ on the potential term is the Kac factor that can be interpreted as a change of time unit valid for any finite $N$,
such that the total energy is extensive. 
The magnetization vector and its components are defined by
\begin{equation}
	\mathbf{M} = \frac{1}{N}\sum_{i=1}^{N}\left(\cos \theta_{i}, \sin \theta_{i} \right)\equiv\left(M_{x},M_{y}\right)=(M\cos\varphi,M\sin\varphi).
	\label{magnt}
\end{equation}

Given the Hamiltonian (\ref{eq:HMF}),
the total internal energy per-particle can be written as
\begin{equation}
  e = \frac{H}{N} = \frac{K}{N} + \frac{1-M^{2}}{2}.
\end{equation}
The motion of particle $i$, generated by the $N$-body self-consistent dynamics of Hamiltonian~(\ref{eq:HMF}), is also the motion of a \emph{test}
particle generated from the ``energy''
\begin{eqnarray}
  U_i 
  & = & \frac{ p^{2}_i }{2 m_i} + 1 - M\cos{(\theta_i - \varphi)} 
  \nonumber \\
  & = & \frac{ p^{2}_i }{2 m_i} + 1 - M_x \cos \theta_i - M_y \sin \theta_i  .
\end{eqnarray}
This single-particle dynamics does not conserve $U_i$, and one easily sees that $\sum_i U_i = 2 Ne - K \neq Ne$. 
This test-particle dynamics provides deep insight in the system evolution \cite{PLP,lhmf_2019} 
because, when $\bf M$ evolves slowly, one may interpret the motion of each particle as that of a pendulum. 
Then one may define a separatrix in the single-particle $(p, \theta)$ space, with energy $U_{\mathrm{s}} = 1 + M$,
delineating particles with $U_i < U_{\mathrm{s}}$ (which exhibit bounded oscillatory motion in the cat's eye) from particles with
$U_i > U_{\mathrm{s}}$ (which travel over the whole circle).
The existence of such a separatrix is the source of instabilities and is related to the chaotic behavior of the system,
as discussed in~\cite{lhmf_2019,latora_2000,Ginelli_2011}.

One usually considers the case of identical particles of unit mass. 
As discussed above, and in order to consider the effect of hard collisions on the dynamics of the system non-trivially, one has
to handle the case with particles of different masses, which is our main goal here.

Note that the canonical ensemble description of this model is almost the same as for a single species. 
Indeed, at temperature $T$ (setting the Boltzmann constant to unity), the partition function is $Z = Z_K Z_V$, where 
\begin{equation}
  Z_K = \int_{\RR^N} \prod_{i=1}^N \rme^{- p_i^2 / (2 m_i T)} \rmd p_i 
          = (2 \pi T)^{N/2} \prod_i \sqrt{m_i}
\end{equation}
and $Z_V$ is exactly the expression given with a single species. Thus, compared with a single-species model with mass $m_1$, 
the free energy of the model with masses $m_1, \ldots m_N$ is 
\begin{equation}
  F(T, m_1, \ldots m_N ) = F(T, m_1, \ldots m_1 ) -  \frac{T}{2}  \sum_i \ln \frac{m_i}{m_1}    .
\end{equation}
This leads to the same internal energy $N \langle e \rangle$.
Ferromagnetic phase transition occurs at the same temperature 3/4 and the same energy per-particle 1/2 as for a single mass species.
Note that the preservation of positional order by the dynamics does not affect the equation of state.

\section{Integration algorithm}
\label{secMD}

The particles are ordered in the initial configuration so that
\begin{equation}
	\theta_1<\theta_2<\cdots<\theta_N.
	\label{orderdef}
\end{equation}
Due to the hard-core collision term, this ordering is
preserved by the dynamics (modulo the crossing at the boundaries at $\theta=0$ and $\theta=2\pi$).
In usual event-driven simulations, the time for the next collision must be computed and all particles are advanced to this time to
implement the collision~\cite{EV}.
Of course, this is not possible for the HMF model as the particle movement between collisions is not integrable.

The collisional time $t_{i,i+1}$ for the next collision between particles $i$ and $i+1$
(considering that the neighbor of particle $N$ at its right is particle $1$)
is computed, for all particles at the beginning of each time step, by supposing that
$F_{i+1}=F_i$, with $F_i(\theta(t))\equiv F_i(\theta_1(t),\ldots,\theta_N(t))$ the
force on particle $i$ due to all other particles, are constant up to the collision time. This yields:
\begin{eqnarray}
t_{i,i+1} &=& -\frac{\theta_{i,i+1}}{v_{i,i+1}},
        \label{tijchmf}
\end{eqnarray}
with $\theta_{i,i+1}\equiv\theta_{i}-\theta_{i+1}$, $v_{i,i+1}=v_i-v_{i+1}$ and $v_i=p_i/m_i$ the velocity of particle $i$.
If the time step $\Delta t$ is small and the interparticle distance
at time $t$ between the particles that will actually collide is small (this condition is met provided $N$ is not too small),
then the error incurred is negligible. Since the array of values $t_{i,i+i+1}$ is updated at each time step, the collisional time
for the next collision will satisfy these conditions, with an ensuing error one
order of magnitude lower than the error in the integration method.

Let us illustrate our approach using a second-order synchronized leap-frog scheme. For the HMF model without hard-core collisions, the algorithm
has the following steps:
\begin{enumerate}
	\item $p_i(t+\Delta t/2)=p_i(t)+F_i(\theta(t))\Delta t/2$,
	\item $\theta_i(t+\Delta t)=\theta_i(t)+[p_i(t+\Delta t/2)/m_i] \Delta t$,
	\item $p_i(t+\Delta t)=p_i(t+\Delta t/2)+F_i(\theta(t+\Delta t))\Delta t/2$.
\end{enumerate}
Collisions occurring in the time interval $(t,t+\Delta t)$ are implemented by subdividing step (2) above as:
\begin{enumerate}[label=\roman*]
	\item Compute for the current value of $t$ all future collision times.
	\item Locate the next collision time $t_{\rm next}$ satisfying $t<t_{\rm next}<t+\Delta t$,
		and the corresponding pair of particles $(i,i+1)$.
	\item Evolve all particles up to time $t_{\rm next}$ as a free motion with velocity $p_i(t+\Delta t/2)/m_i$
		and implement the point-like collision between particle $i$ and $i+1$.
	\item Repeat steps (ii) and (iii) until there are no collisions to implement in the time interval $(t,t+\Delta t)$.
\end{enumerate}
The spatial ordering of the particles is kept by the exact dynamics, but small errors in the integration scheme may lead to some particle
being in a wrong ordering. As a consequence, after each time step a fast sequential search, with computer effort proportional to $N$,
is performed. This ensures that particles are always ordered according to Eq.~(\ref{orderdef}).

\section{Results}
\label{sec.rslts}

As explained in the introduction, hard-core collisions in one dimension for identical particles do not change the time
evolution of the system statistics. We thus consider particles with two different masses $m_1=1$ and $m_2=5$ with
a proportion $\nu_1=N_1/N$, where $N_1$ is the number of particles with masses $m_1$ and $N$ is the total number of particles.
It is worth noticing that, for the usual HMF model, the dynamics naturally leads to mass segregation~\cite{zolacir}, while here
no mass segregation is possible as the ordering in space of different masses is conserved due to the hard-core collisional force.
In all simulations presented here, the initial state is a waterbag state (identical for both species), a uniform distribution in the intervals
$-\theta_0<\theta<\theta_0$ and $-p_0<p<p_0$, where $\theta_0$ and $p_0$ are obtained from the chosen initial magnetization
and total energy.

\begin{figure}[htb!]
\centering
\includegraphics[scale=0.27]{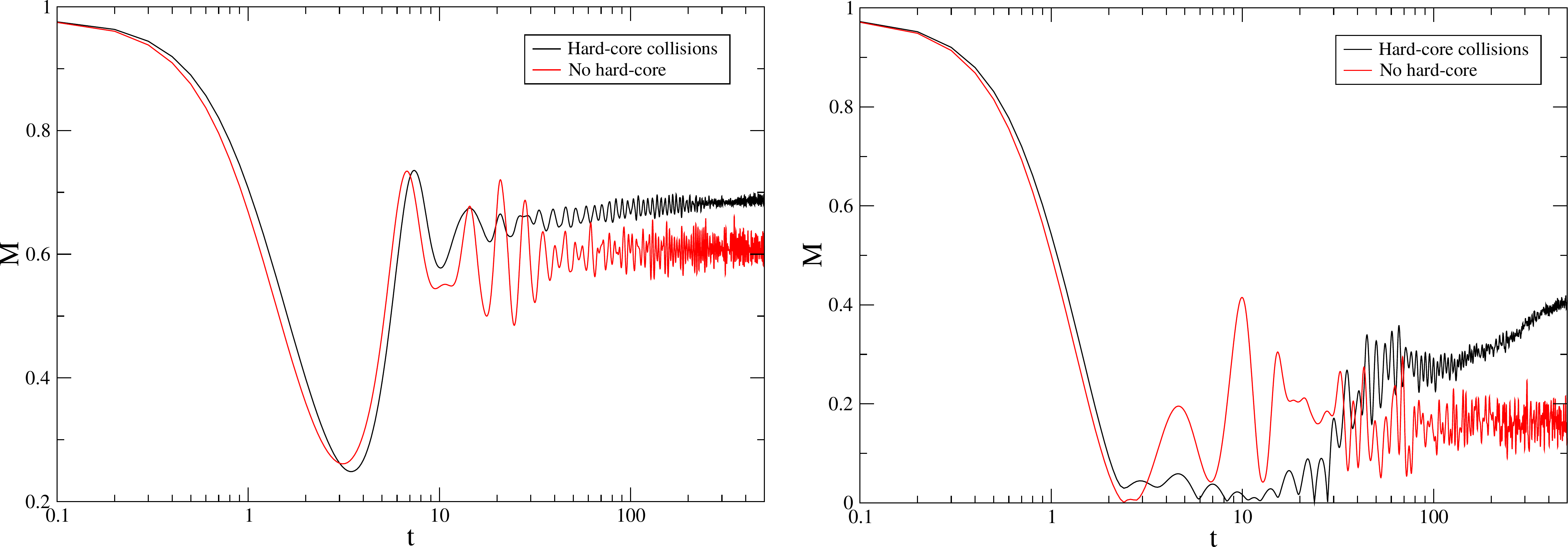}
\caption{Left panel: Magnetization as a function of time for the duration of the violent relaxation for $e=0.5$, $N=10000$,
	initial magnetization $M=0.98$, and averaged over 10 realizations, with and without a hard-core in the interparticle
	interaction potential. Right panel: same as the left panel but with $e=0.8$.}
\label{magcomp}
\end{figure}

\begin{figure}[htb!]
\centering
\includegraphics[scale=0.27]{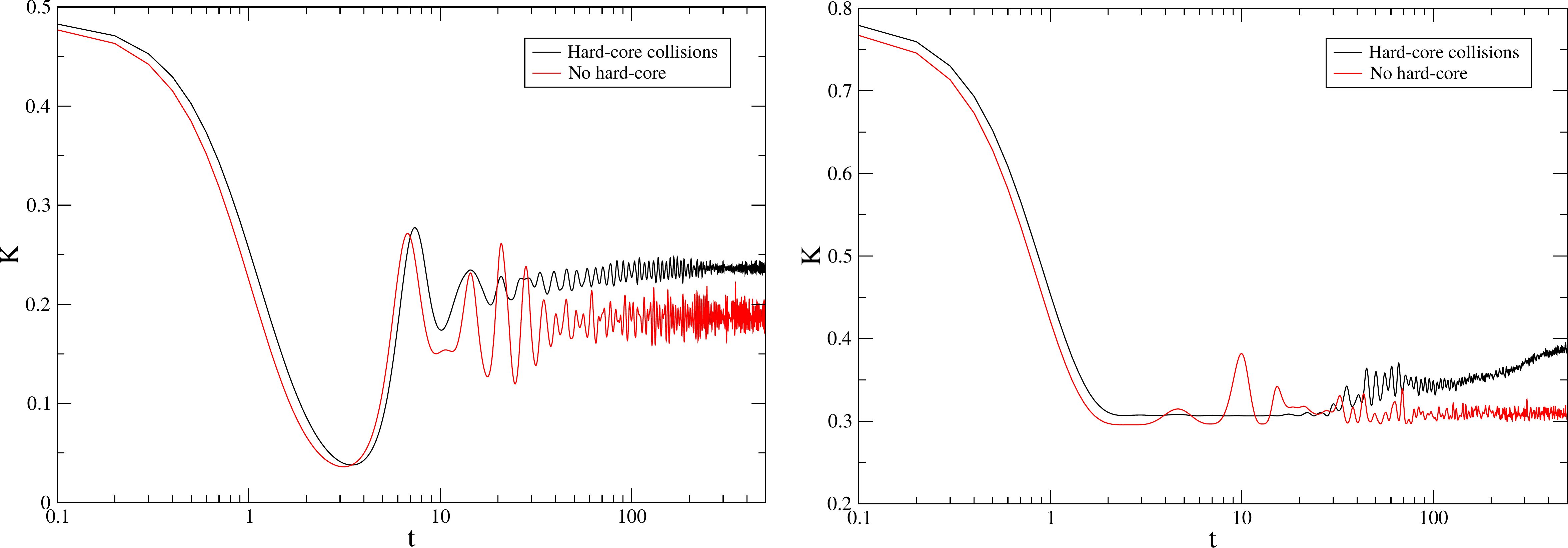}
\caption{Left panel: Kinetic energy as a function of time for the duration of the violent relaxation for $e=0.5$, $N=10000$,
	initial magnetization $M=0.98$, and averaged over 10 realizations.
	Right panel: same as the left panel but with $e=0.8$.}
\label{ener}
\end{figure}

Figures~\ref{magcomp} and~\ref{ener} show the total magnetization $M$ and kinetic energy $K$ as a function of time for
$\nu_1=0.9$ and an interval encompassing the
initial violent relaxation, for $N=10000$ particles and two different initial energies per particle $e=0.5$ and $e=0.8$, for both the HMF
model with and without hard-core collisions.
A first important observation is that the QSS resulting from the violent relaxation is different in both cases, having different magnetizations.
For the hard-core collisional case, the state resulting from the violent relaxation starts to change much more rapidly than for the case without
a hard-core term in the potential. This can be explained from kinetic theory by the fact
that collisional corrections to the Vlasov equation comprise two contributions
for the case with a hard-core term in the interactions:
a Balescu-Lenard type, essentially due to finite $N$ granularity effects, and a Boltzmann like collisional integral.
The former is of order $1/N$ or $1/N^2$ for a non-homogeneous or
homogeneous state, respectively~\cite{scaling,scaling2,fouvry}, while the Boltzmann term is $N$-independent~\cite{balescu}. As a consequence, although
apparently small, the collisional contributions from the hard-core potential to the kinetic equations dominate the system dynamics.
The determination of an explicit kinetic equation for the present case is beyond the scope of this work and will be the subject
of a forthcoming paper.

\begin{figure}[htb!]
\centering
\includegraphics[scale=0.3]{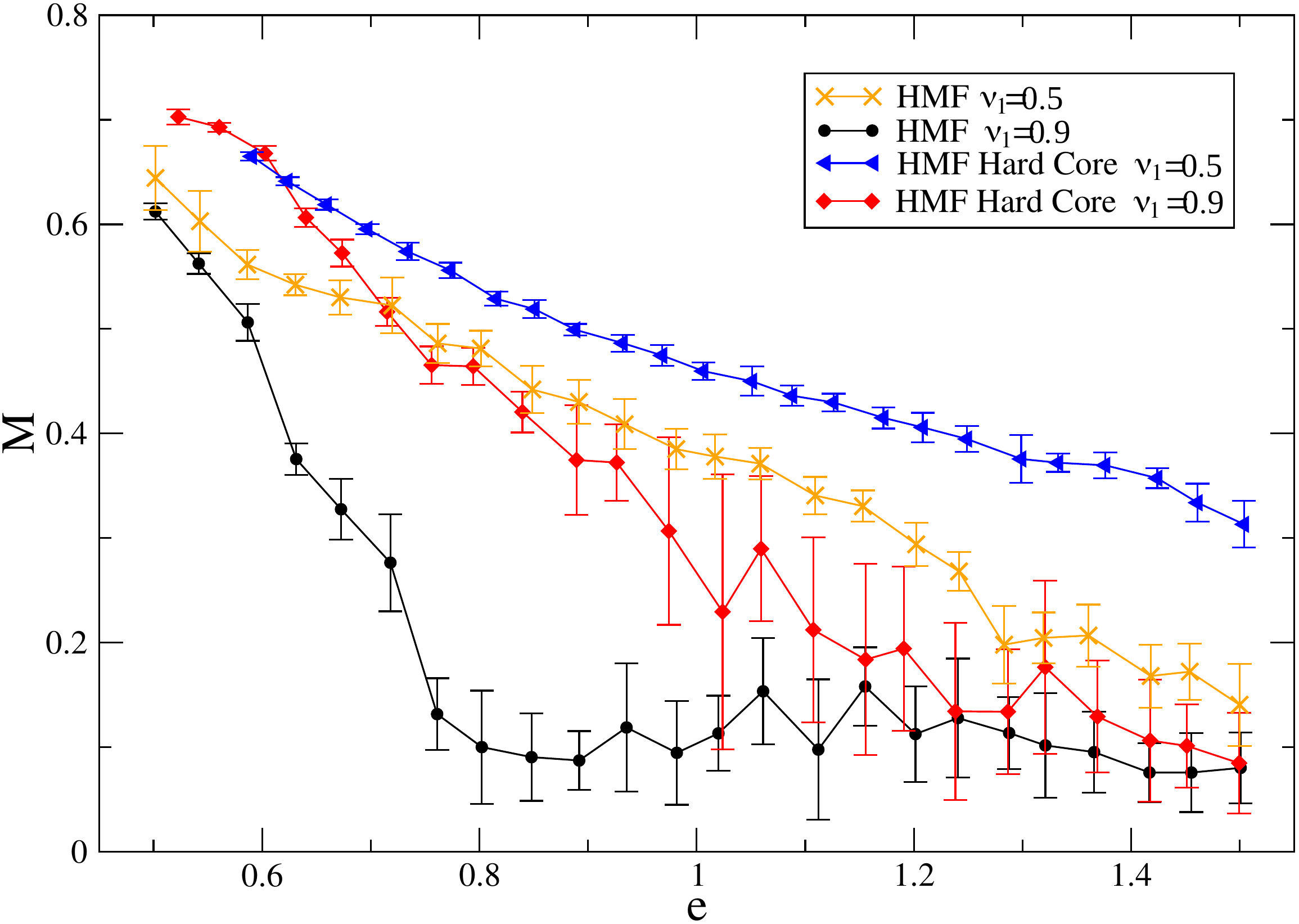}
	\caption{Magnetization of the QSS resulting from the violent relaxation of the initial waterbag state for $N=10000$
	and integration time of $t_{\mathrm{f}}=500$ for the HMF model with and without hard-core collisions and two values of $\nu_1=N_1/N$.
	The error bars were obtained from the standard deviation over the realizations.
	All states have the same initial magnetization value of $M=0.98$.}
        \label{phasediag}
\end{figure}

Figure~\ref{phasediag} shows the magnetization as a function of energy after the violent relaxation at time $t_{\mathrm{f}}=500$ for a few energy values,
and two different proportions $\nu_1=0.5$ and $\nu_1=0.9$ for both cases (with and without a hard-core).
It becomes again evident that the QSS after the violent relaxation is changed by the presence of hard-core collisions. For higher energies, the
system becomes homogeneous, as can be easily seen from molecular dynamics simulations (not shown here).
Figure~\ref{phasediag} illustrates the fact that the critical energy per-particle, for the transition from non-homogeneous to homogeneous states,
is lower the higher the value of the mass proportion $\nu_1$.
This is expected since, for a given energy per-particle $e$, the bigger
the average mass, the smaller the velocities. Intuitively we  see that this must be case by noting that one can raise the numerical value
of the average mass in the system by a change of time unit, and consequently also of the unit of energy.

\begin{figure}[htb!]
\centering
\includegraphics[scale=0.3]{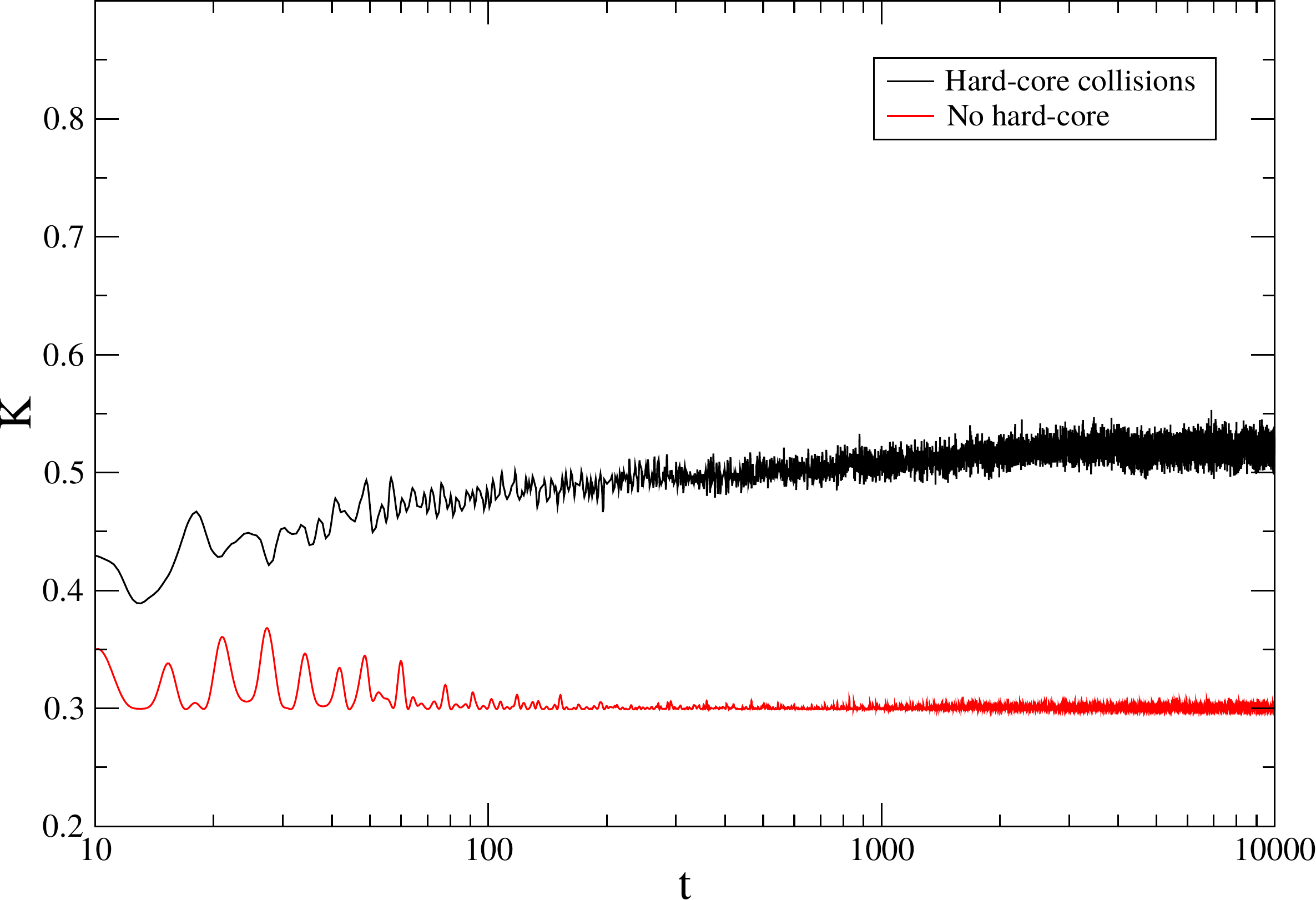}
	\caption{Time evolution of the kinetic energy for $e=0.8$, $N=1000$, $M(t=0)=0.98$ for HMF model with and
	without a hard-core potential, $\nu_1=0.9$, up to the final equilibrium state.}
        \label{longtime}
\end{figure}

We note also that the existence of a hard-core force at zero distance changes the final equilibrium state for a given energy. Figure~\ref{longtime}
shows the time evolution of the kinetic energy up to the final simulation time of $t_{\mathrm{f}}=10000$ for $N=1000$ for both the collisional and non-collisional cases
for $e=0.8$, where it is evident that the system tends to states with different final effective temperatures
($T=2K$) for the same energy,
which at its turn implies that the final magnetization is different. This can be explained physically,
at least in part, by the fact that no mass segregation is possible
for the collisional case, and the distributions of the two types of particles remain uniform (the initial distribution was chosen this way), while
in the non-collisional case mass segregation occurs. This leads to different mean-fields and therefore to different equilibrium states.
It would be interesting to investigate a two or three-dimensional system, where mass segregation is not
hindered by the presence of a hard-core, to verify whether the QSS resulting from the violent relaxation changes if a hard-core is added to the potential.

\section{Concluding remarks and perspectives}
\label{sec.cncl}

We introduced the effects of a hard-core point-like (at zero distance) interaction into the HMF model,
and proposed a numerical approach to integrate numerically this model. The method was implemented using a second-order leap-frog scheme
alongside an event-driven method in the spatial evolution part of the leap-frog.
The same approach can be straightforwardly generalized to
other numerical schemes, such as the fourth order Runge-Kutta and symplectic integrators.
We showed that the introduction of a hard collision interaction changes the outcome of the violent
relaxation in the HMF model, and also changes strongly the dynamics of the QSS, resulting in different equilibrium states
for the same energy, depending on whether a hard-core is present or not.
The extension to higher dimensions is also possible and is the object of ongoing research.
It is particularly relevant to investigate whether the QSS after the violent relaxation is affected by the
hard-core term, with equal or different particle masses, for more than one spatial dimension.
This phenomenology may be important in situations where a short distance
strong interaction is present, due for instance to a strong repulsive (or attractive) Coulombian interaction, 
or in a granular medium.

\section*{Acknowledgments}

I.~M.\ thanks CAPES (Brazil) for financing his stay at Aix-Marseille Universit\'e,
through the Programa de Doutorado Sandu{\'\i}che no Exterior (PDSE), grant No. 88881.131521/2016-01.
T.~M.~R.~F.\ thanks CNPq (Brazil) for partially funding this project through the grant No.\ 305842/2017-0.
L.~H.~M.~F.\ thanks CNPq and FACEPE for funding from grant No.\ APQ-0198-1.05/14.
A.~F.\ thanks CNPq for partially funding this project from grant No.\ 307192/2016-4. 
I.~M.\ and Y.~E.\ thank members of Equipe Turbulence Plasma
at Aix-Marseille Universit\'e for many fruitful discussions.

\end{document}